\documentclass[aps,prd,onecolumn,groupedaddress,showpacs,nofootinbib,amssymb]{revtex4-1}
\usepackage[T1]{fontenc}
\usepackage[latin1]{inputenc}
\usepackage{graphicx}
\usepackage[english]{babel}
\usepackage{graphicx}
\usepackage{bm}
\usepackage{amsmath}
\usepackage{amssymb}
\usepackage{amsfonts}
\usepackage{epsfig}
\usepackage{colordvi}
\usepackage{color}

\begin{document}
%\usepackage{your_package}	%if you need custom package
%\usepackage[notquote]{hanging}

% * CHAPTER NUMBER * BOOK NAME * AUTHOR(S) NAME *****************************
%\setcounter{chapter}{0} % It will be set by technical editor.

%\booktitle{Will-be-set-by-IN-TECH}%

%\chaptertitle{The missing matter problem: from the dark matter search to alternative hypotheses} % You know your chapter title?

\title{The missing matter problem: from the dark matter search to alternative hypotheses}

\author{S. Capozziello, L. Consiglio, M. De Laurentis, G. De Rosa, C. Di Donato}

\affiliation{Dipartimento di Scienze Fisiche, Università
di Napoli {}``Federico II'', Compl. Univ. di
Monte S. Angelo, Edificio G, Via Cinthia, I-80126, Napoli, Italy\\
INFN Sezione  di Napoli, Compl. Univ. di
Monte S. Angelo, Edificio G, Via Cinthia, I-80126, Napoli, Italy.}

\begin{abstract}
Dark matter is among the most important open problems in both astrophysics and particle physics. 
Recently, hints of anomalous cosmic-ray spectra found by astroparticle experiments, such as PAMELA, have motivated interesting interpretations in terms of DM annihilation and/or decay. Even if these signatures also have standard astrophysical interpretations, Fermi-LAT electron spectral measurements indicated the presence of an additional positron source, which could be DM annihilation or decay. Searches by the Fermi-LAT for gamma-ray signals have also been performed, along with measurements of the diffuse Galactic and extragalactic gamma-ray emission, providing unprecedented high quality data and statistics which makes crucial to investigate DM in gamma rays. In addition, Imaging Air Cherenkov Telescopes like HESS, MAGIC, and VERITAS have reported on searches for gamma-ray emission from dwarf galaxies.
Moreover, CMS and ATLAS experiments at the Large Hadron Collider (LHC) currently in operation at CERN, are giving their first results for supersymmetry searches and for discovering DM in colliders.
Concerning direct search of DM, the most stringent limits on the elastic spin-independent WIMP-nucleon cross-section are coming from CDMS-II, EDELWEISS-II and , in particular, XENON100, whereas, recently, the CoGeNT collaboration reported the WIMP candidate signal events exceeding the known backgrounds and results obtained by the DAMA/LIBRA set-up show the model independent annual modulation signature for DM particles.
On the other hand, DM and Dark Energy (DE)  could be nothing else but the signal that General relativity is not working at large scales (infrared scales) and alternative theories of gravity should be considered in order to fit observations. The issues of Òmissing matterÓ and Òaccelerating universeÓ could be addressed by taking account extensions of General Relativity (e.g. Extended Theories of Gravity) where gravitational interaction works in different ways at different scales. From this point of view,  the gravitational sector should be revised without invoking the presence of new ingredients that, up to now, have not been detected at a fundamental level.  At ultraviolet scales,  the production of massive gravitons would be a test-bed for the theory.
This review presents many aspects, from astrophysical observations to particle physics candidates and describes the theoretical and experimental aspects of the DM problem (or missing matter, considering the alternative approaches)  in particle physics, astrophysics and cosmology. 
A brief overview is given of the phenomenology of several dark matter candidates and their expected production mechanisms, basic properties, and implications for direct and indirect detection, particle colliders, and astrophysical observations.
A summary of the experimental status is given and the possible scenarios opening with the upcoming are discussed in the context of an approach which combines information from high-energy particle physics with cosmic-ray and traditional astronomical data.
\end{abstract}

\maketitle

\section{Introduction}

The nature of dark matter (DM) is one of the greatest today challenges
in cosmology and particle physics: on one side it would have a
significant impact on the large scale structure in the Universe and,
on the other hand, it should lead to the empirical evidence of new
unknown particles. The difficulties arise from the fact that DM could
be composed by multiple components behaving in different ways
depending on the scale. Besides, DM dynamics should be connected to
that of dark energy (DE), the other unknown, unclustered form of
energy that recent observations pointed out in the last decade. This
further ingredient should constitute almost 75 \% of cosmic
matter-energy budget, turning out that the so-called {\it "dark side"
}problem is dramatic and urgent to be solved.

The  DM problem was finally set in the 1970s after several evidences accumulated in the past decades. In 1933 Fritz Zwicky measured
the velocity dispersion of galaxies in the Coma cluster and found out that it was about a factor ten larger than expected from the estimated total mass of the cluster
\citep{Zwicky}. It is interesting that Zwicky defined such a shortcoming as {\it the missing matter problem}. These preliminary  observations have been confirmed in  the 1970s, when  data collected on the galactic rotational curves of spiral galaxies, proved the presence of large amounts of mass on scales much larger than the optical size of galactic disks. 

Later on,  the evidence of DM at various cosmological and astrophysical scales  has been established
by a wide number of  observations, especially the
very precise measurements of the cosmic microwave background radiation
in the Wilkinson Microwave Anisotropy Probe (WMAP) experiment
\citep{WMAP}. Data from weak \citep{weaklensing} and strong
\citep{stronglensing} lensing, hot gas in clusters \citep{Chandra}, the
Bullet Cluster \citep{Clowe}, Big Bang nucleosynthesis (BBN)
\citep{PDG-BNN}, further constraints from large scale structure
\citep{X-ray}, distant supernovae \citep{SN-1, SN-2}, and the cosmic
microwave background (CMB) \citep{WMAP-2} also support the evidence for
non-luminous matter.

The ensemble of these data provides a strong evidence for a non
luminous and non absorbing matter   interacting only gravitationally,
which is five times more prevalent than ordinary matter and accounts
for about a quarter of the Universe. More precisely, current data
constrain the energy densities of the Universe in a baryonic component,
mostly known ($\Omega_B \lesssim 0.0456 \pm 0.0016$), a DM (CDM)
component ($\Omega_{CDM} \simeq 0.227 \pm 0.014$) still unknown, and a
DE component ($\Omega_{\Lambda} \simeq 0.728 \pm 0.015$
with a great uncertainty on the generating mechanisms). The luminous
matter in the Universe is less than 1$\%$ of the total composition of
the Universe. The current most precise estimation of the density of
non-baryonic DM $\Omega_{DM}$ is obtained combining the measurements
of the CMB anisotropy and of the spatial distribution of the galaxies
and has found to be $\Omega_{DM} h^2 = 0.110 \pm 0.006$. The
"local" DM present in the Galactic disk has an average density of
\citep{LocalDensityPDG}: $\rho_{DM}^{local} \simeq 0.3
\frac{GeV}{cm^3}$.
 
An alternative, intriguing approach is to consider DM, as well as DE, as the
manifestation of the break-down of General Relativity (GR) on large
scales. A large literature is devoted to the possible modifications of the
laws of gravitation, as Extended Theories of Gravity (ETGs), in
particular $f(R)$-theories, by introducing into the Lagrangian ,
physically motivated higher-order curvature invariants and
non-minimally coupled scalar fields \citep{reportnoi}. The interest on
such an approach, in early epoch cosmology, is due to the fact that it
can ``naturally'' reproduce inflationary behaviours able to overcome
the shortcomings of  Standard Cosmological Model and, in late epoch cosmology, it  seems 
capable of matching with several observations overcoming  DM and DE problems related to the issue of detection.

In this review paper, we illustrate the experimental status of art to detect DM
particles with direct and indirect methods, trying to interpretate and
discuss the results obtained so far (Secs. 2,3,4).  A possible alternative approach to the 
 DM problem is presented in terms of gravitational effects
at  astrophysical and cosmological scales (Sec.5). Specifically,  corrected gravitational potentials
 offer the possibility to fit galaxy rotation curves and galaxy clusters haloes without  DM. The same approach allows to fit the apparent accelerated  Hubble fluid without invoking DE.  Conclusions are drawn in Sec. 6.
%%%%%%%%%%%%%%%%%%%%%%%%%%%%
\section{A survey of Dark Matter candidates}
%%%%%%%%%%%%%%%%%%%%%%%%%%%%%
%%Analyses of the structure formation in the Universe [\citep{inserire Ref
%%PDG}] suggest that most DM should be "cold", i.e.,
%%non-relativistic at the onset of galaxy formation.

Candidates for non-baryonic DM must satisfy several conditions: $i)$ they
should be neutral, (otherwise they would interact
electromagnetically);  $ii)$ they should not have color charge (otherwise
they could form anomalous nuclear states);  $iii)$ they should be stable on
cosmological time scales (otherwise they would have decayed by now); $iv)$
they should interact very weakly with ordinary matter (otherwise they
would not  be dark), and, finally, $v)$  they should have a suitable  relic density. 
If we consider the Standard Model (SM) of  Particles, neutrinos seem to be
the prominent DM candidate as they interact only weakly and have an
extremely low mass (the exact value of neutrinos' mass has yet to be
measured, but it is clear, from neutrino oscillation measurements,  that their mass is
non-zero). There are many sources of neutrinos in the Universe,
nevertheless they can account only a small fraction of DM. First of
all,  they are "hot" since they move at relativistic velocities. Thus
they should have had  a strong effect on the instabilities that 
generated the primordial  cosmological objects during the earliest Universe,
in the sense that galaxy clusters would have formed before galaxies
and stars. This is in contrast with most theories and measurements,
which are, instead, supported by a model based on cold DM, consisting of
particles which move at non-relativistic energies.  Thus small-scale
perturbations are not suppressed, allowing an early start of 
structure formation. Secondly, we
know that neutrino mass is rather small. The most recent upper bound
on electron neutrino mass is $m_{\nu_{e}} <$ 2 eV (95\%
c.l.)  while the experimental limits on the muon and tau
neutrino are even weaker. So neutrinos cannot explain the 
gravitational effects DM is responsible for. However, extremely, high
energetic neutrinos are of interest for DM search as they are among
the secondary particles created in the annihilation of other DM
candidates. 

The most popular hypothesis is that non-baryonic DM
consists of some neutral massive weakly interacting particles (WIMPs),
which were created in the hot early Universe, decoupled early from
ordinary matter in order to seed structure formation, and survived
until today.
They are supposed (from theoretical considerations and from the fact
that they have not been detected yet) to be heavy respect to SM
particles, with mass roughly between 10 GeV and few TeV. They
interact via weak force and gravity only and  are stable with a
lifetime at least equal to the age of the Universe. Their relic
density can be correctly calculated assuming that WIMPs were in
thermal equilibrium with the SM particles in the early Universe.
At the early stages, Universe was dense and hot. As the temperature  $T$
of the Universe cools, the density of more massive DM particles, with
mass greater than $T$, become Boltzmann exponentially suppressed. When
the expansion rate of the Universe, $H$, exceeds the particle
annihilation/creation rate, the WIMPs drop out of thermal equilibrium,
and the number density becomes "frozen". Presently the relic density
of these particles is approximately  given by
 
\begin{equation}
\frac{\Omega_{DM} h^2}{0.110} \approx \frac{3 \times 10^{-26} cm^3 /sec} {\langle \sigma_{A} v \rangle_{ann}} ,
\end{equation}

where $\sigma_{A}$ is the total annihilation cross section, $v$ is the
relative velocity of WIMPs and the term in brackets is an average over
the thermal distribution of WIMPs velocities. From a cosmological point
of view, it is  worth noticing that the proper value of
$\Omega_{DM} h^2$ density comes out from an annihilation cross section
on the electroweak scale. This coincidence, obtained numerically,
represents the main reason for believing that WIMPs give the largest
contribution to the matter density  in the Universe.

The freeze out happens at temperature $T_F \simeq m_{DM}/20$ almost
independently of the properties of the WIMPs. This means they are
already non-relativistic at the decoupling.

If we search beyond SM, well-motivated scenarios
suggest good candidates. As neutralinos and Kaluza-Klein particles on
which we are going to focus our discussion.

%%%%%%%%%%%%%%%%%%%%%%%%%%%%%%
\subsection{WIMPs in supersymmetric extensions of the Standard Model}
%%%%%%%%%%%%%%%%%%%%%%%%%%%

This class contains a large amount of DM candidates, which are not
predicted in the realm of the SM. 
In supersymmetric (SUSY) theories, each SM particle has a new
yet-undiscovered partner whose spin differs by 1/2 with respect to the
supersymmetric partner. In comparison to the mass eigenvalues in the
SM, SUSY particles occur in linear combinations. The
lightest possible combination (LSP) is the neutralino $\chi$ formed by
the Bino \~{B}, Wino \~{W} and two Higgsino \~{H} states, with a mass
range $m_{\chi}$ $\sim$ 10 GeV-TeV. The main ingredient necessary to
provide a natural WIMP candidate in the SUSY models is the
R-parity conservation defining the R-parity as:

\begin{equation*}
R=(-1)^{3(B-L)+2S}\,,
\end{equation*}

where B is the baryon number, L the lepton number and S the spin of
the particle. $R = +1$ for ordinary particles and $R = -1$ for
SUSY particles. This means that SUSY particles can
only be created or annihilated in pairs. By imposing the R-parity
conservation, a selection rule on the SUSY particle decays
prevents the LSP to decay to an ordinary particle guaranteeing the
stability in terms of cosmological abundances. However, this straightforward mechanism has to be experimentally probed. 
Several indications at Large Hadron Collider (LHC), CERN,  seem to exclude minimal SUSY models, so the search of these candidates is, up to now, completely open.

%%%%%%%%%%%%%%%%%%%%%%%%%%%%%%%%%%%%%% 
 \subsection{Extra dimensions and Kaluza-Klein Dark Matter}
%%%%%%%%%%%%%%%%%%%%%%%%%%%%%%%%%%%%%%%

An alternative possibility for new weak-scale physics is to search for universal
extra dimensions (UED). The motivation to consider UED models is that
they provide an interesting and qualitatively different alternative to
supersymmetry. The idea of the existence of extra spatial dimensions
was introduced, for  the first time, by Kaluza and Klein in the
1920's. Such theories attempt to unify the two fundamental forces of
gravitation and electromagnetism at the weak scales. Some models assume
that all  fields of  SM propagate in ``universal''
extradimension. As a consequence, observers in the
four-dimensional world see a tower of Kaluza Klein (KK) states for
each SM particle \citep{Kolb}.  The SM particles make up the first level of this
tower and are referred to as the zero-th KK mode.

Momentum conservation in the extra dimension leads, in the four
dimensional world, to a conservation of KK number ($N_{KK}$), where
the $KK$ number of a particle is given by its mode number. All SM
particles have $N_{KK}$ = 0 while the next most massive set of states
have $N_{KK}$ = 1. Such KK excitations appear as particles with
masses near the TeV scale. Due to conservation of momentum in the
higher dimensions, a symmetry called KK parity can arise which can, in
some cases, make the lightest KK particle (LKP) stable, in a way
which is analogous to R-parity of SUSY models, making it
possible for the LKP to be a viable DM candidate.
Since DM particles are expected to be neutral, non-baryonic
and without color, the first mode KK partners of the neutral gauge
bosons or neutrinos are likely choices for the lightest
Kaluza-Klein particle (LKP).
The identity of the LKP depends on the mass spectrum of the first KK
level. The LKP is, most naturally, the first KK excitation of the
$B_1$, the level 1 partner of the hypercharge gauge boson. It has been
found that the appropriate relic density is predicted when the mass is
moderately heavy, between 600 and 1200 GeV, somewhat heavier than
the range favoured by supersymmetry. This range of  LKP mass
depends on the details of the co-annihilations of LKPs with heavier KK
particles. Another difference between DM particles in universal extra
dimensions and supersymmetry is that, unlike of LSP neutralino, the
bosonic nature of the LKP means there is no chirality suppression of
the annihilation signal in fermions. The annihilation rate of  the LKP
is therefore roughly proportional to the hypercharge of the final
state, leading to a large rate in leptons, including neutrinos. The
annihilation and co-annihilation cross-sections are determined by
SM couplings and the mass spectrum of the first KK level.
In contrast to supersymmetry,  particles in UED have the same spin
of the  SM partners. As a result, the couplings become large and
non-perturbative at energies far below the Planck scale. The detection of these kind of particles is also expected at LHC.
 
%%%%%%%%%%%%%%%%%%%%%%%%%%%%%%%%
\section{Methods of Detection and experimental status}
%%%%%%%%%%%%%%%%%%%%%%%%%%%%%%%

There are three main strategies to search for DM particles. Assuming that
DM consists of WIMPs i.e. particles which froze out from thermal
equilibrium when they were no more relativistic, one would expect that
due to the gravitational interaction, they should have clustered with
ordinary matter to form an almost spherical halo around the
galaxies. Measurements of rotation curves of a large number of spiral
galaxies have suggested the existence of a dark halo more extended
than the visible disk. Despite the measurement in case of Milky Way is
more difficult and suffers of larger uncertainties, the presence of a
halo has been confirmed \citep{merrifield}.
Thus, if DM exists, it should be present also in the Milky Way and a
consistent flux of WIMPs is expected to cross the Earth surface. The
WIMPs of the dark halo, although with a very low cross section,
interact with ordinary matter inducing atomic recoils. The measurement
of these rare recoils through their energy deposit, carefully
discriminated from the background, is performed by direct techniques.

Another method for DM detection is carried out by indirect techniques, which
aims to observe the products of WIMPs annihilation process, as gamma
rays, neutrinos, anti-matter cosmic rays occurring mainly inside
astrophysical objects. 

A third possibility is at
particle colliders: here DM may be produced through the SM SM
$\rightarrow$ XX process where SM denotes a standard model particle and X
represents the WIMP. Such events are, in general, undetectable but are typically
accompanied by related production mechanisms, such as SM SM
$\rightarrow$ XX + ``{SM}'' , where ``{SM}'' indicates one or more standard
model particles. These events are instead observable and should provide
signatures of DM at colliders as LHC.

%%%%%%%%%%%%%%%%%%%%%%%%%%%%%
\subsection{WIMPs Direct Detection}
%%%%%%%%%%%%%%%%%%%%%%%%%%%%%%%%

When a WIMP of a certain mass $m_{X}$ scatters elastically a
nucleus of mass $m_{N}$, the nuclear recoil occurs at an angle
$\theta$ with respect to the WIMP initial velocity, with $\cos
\theta$ uniformly distributed between -1 and 1 for the isotropic
scattering that occurs with zero-momentum transfer. If the WIMP
initial energy in the lab frame is  $E_i = M_X v^2/2$, the
the nucleus recoils (in the lab frame) with energy
$\displaystyle{E_R = E_i r \frac{(1-\cos \theta)}{2}}$
where
$\displaystyle{r \equiv \frac{4 \mu^2}{M_X M_A} = \frac{4 M_X M_A}{(M_X + M_A)^2}}$.
Note that $r \leq 1$, with $r = 1$ only if $M_X = M_A$. For this
isotropic scattering, the recoil energy is therefore uniformly
distributed between $0-E_i r$.

Since typically the mean WIMP velocity is $v \simeq 220 km/s$ \citep{Berna}, we can
easily estimate that the maximum energy transfer from a WIMP to an
electron, initially at rest, is at most in the eV range, while the
energy transfer to an atomic nucleus would typically be in the range
of tens of keV. This requires to use low threshold detectors, which
are sensitive to individual energy deposits of this order of
magnitude. In order to compute the WIMP-nucleon interaction rate, one
needs the cross-section and the local density of WIMP. Details of this
calculations can be found in e.g. \citep{jungman,lewin}. We report here only the final expression

\begin{equation*}
\frac{dN}{dE_R} =
\frac{\sigma_0 \rho_{DM}^{local}}{2\mu^2 m_X} F^2(q)\int_{v_1}^{v_2} \frac{f(v)}{v}dv
\end{equation*}

\begin{figure}[htb]
\centering\includegraphics[width=70mm]{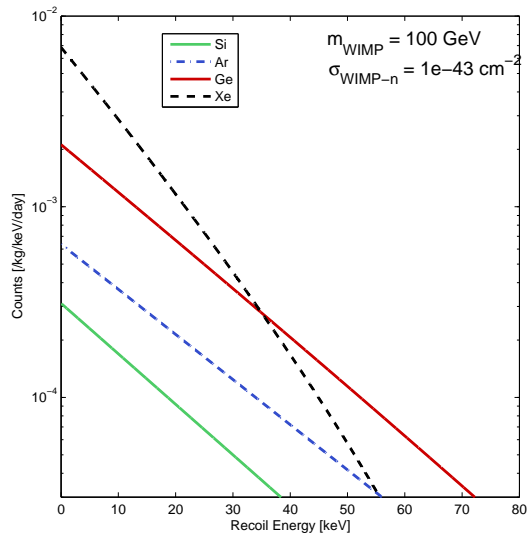}
\caption{Differential WIMP recoil spectrum for a WIMP mass of 100\,GeV 
and a WIMP-nucleon cross section $\sigma = 10^{-43}$ cm$^2$. The spectrum was 
calculated for Si (light solid), Ar (light dot-dashed), Ge (dark solid), Xe (dark dashed) \citep{baudis}.} 
\label{rates}
\end{figure}

\vspace{0.3 cm}

where $\rho_{DM}^{local}$ is the local WIMP density, $\mu$ is the
reduced mass of the WIMP nucleus system and $f(v)$ accounts the
velocity distribution of WIMPs in the galactic halo. The lower and
upper limit of the integral represent respectively the minimum WIMP
velocity to produce a recoil of energy $E_r$, and the maximum WIMP
velocity set by the escape velocity of the halo model. $F(q)$ is a
dimensionless factor form and $\sigma_0$ is the WIMP-nucleus
interaction cross-section.  This cross section depends strongly on the
form of the interactions of the DM particles. It may happen that the
WIMP-nucleus cross-section is not sensitive to the spin of the
nucleus; in such a case we refer to spin-independent interactions;
otherwise, WIMPs couple dominantly to the nucleus spin and this is the
case of a spin-dependent interaction. The WIMP-nucleus cross section,
may be written in terms of a spin-independent (mostly scalar) and a
spin-dependent (mostly axial vector) component. For the former, the
interaction will be coherent across the nucleons in the nucleus, while
the latter term will only be present for nucleons with nuclear spin
29 or 82 or 83). In most cases, the coherent term is dominant
since it is proportional to $A^2$ where A is the atomic number of
nucleus, which favours heavy nuclei. Nevertheless, for very heavy
nuclei, as the recoil energy increases, account must also be taken of
the nuclear form factor, which may suppress the differential
scattering rate significantly as shown in Fig.\ref{rates}. The curves
are obtained assuming the spin independent coupling dominant, a
standard halo model and choosing a WIMP mass of 100 GeV \citep{baudis}.

 The differential WIMP event rate versus the recoil energy expected for single isotope targets of $^{131}$Xe (similar for $^{129}$I), $^{73}$Ge, $^{28}$Si and $^{40}$Ar is shown. It can be
noticed that for a given cross section for WIMP-nucleon interactions, the interaction rate decreases
for smaller nuclei because of a combination of smaller coherence enhancement
($\sim A^2$) and the less effective transfer of recoil energy to a target that is lighter
than the WIMP. It is also evident that the recoil spectrum for the heavier Xe nucleus is 
suppressed significantly by the loss of coherence for higher $q^2$ scattering events (form factor
suppression). A low analysis threshold is therefore important to maximise the effective
search sensitivity of a given detector mass. The influence of threshold
energy is even greater for lower-mass WIMPs, where the recoil spectrum slope
becomes steeper because of the reduction in typical kinetic energy of the WIMPs.
The interaction rate can be calculated within the Minimal
SUSY models, but the predicted WIMP-nucleon interaction
cross section spans many orders of magnitude. Typical values for the
spin-independent cross section are between $10^{-6}$ pb and $10^{-11}$
pb \citep{de Austri et al. 2006}. 

 In the recent years, an inelastic dark matter (iDM) model has
  been proposed as a modification of
  the elastic WIMP model. iDM model assumes that WIMPs scatter off
  baryonic matter by simultaneously transitioning to an excited state
  at an energy $\delta$ above the ground state ($\chi N \rightarrow
  \chi^* N$), while elastic scattering is forbidden or highly
  suppressed. This fact introduces a minimum velocity for WIMPs to scatter
  in a detector with a deposited energy E$_{nr}$.
  
%%%%%%%%%%%%%%%%%%%%%%%
\subsubsection{A Review of Experiments}
%%%%%%%%%%%%%%%%%%%%

The nature of WIMPs makes their detection a hard task. A typical WIMP
mass ranges between 10 GeV and few TeV according to the model chosen and
the signal, to be detected in terms of deposit energy of the nuclear
recoil, is of the order of tens keV.  Within the MSSM models, the
cross section interaction rates induce at most 1 evt day$^{-1}
kg^{-1}$ in the detector, so large target masses and low background
detectors are required. Typical background from environmental
radioactivity and cosmic radiation are much higher with respect to these low
expected rates and experimental purposes need underground laboratories
to shield  cosmic ray induced backgrounds, and for the selection of
extremely radio-pure materials.

Worldwide, a large number of underground laboratories exists, many of
them house present and/or future DM experiments (see Tab. \ref{tab:UL}). 
\begin {table}
\caption{Underground laboratories housing DM experiments. The
approximate effective shielding depth is mesured in meters of water
equivalent (mwe). }
\centering
\begin{tabular}{|l|c|r|} 
\hline 
Laboratory &  Depth (m.w.e.) & Experiment \\
\hline
Soudan, US &   2000 & CDMS/SuperCDMS, CoGeNT \\
\hline 
Yangyang, Korea & 2000 & KIMS \\
\hline
Kamioka, Japan &  2700 & XMASS,SuperKamiomkande \\
\hline
Bulby, UK & 3200 & ZEPLIN \\
\hline
LNGS, Italy & 3500 & DAMA, CRESST, WARP, XENON100 \\
\hline
Modane, France & 4800 & EDELWEISS \\
\hline
SNOLab, Canada & 6000 & PICASSO, DEAP/CLEAN \\
\hline 
\end{tabular}
	\label{tab:UL}
\end{table}

Three different detection principles are at the basis of most particle detectors:
\begin{itemize} 
\item {\bfseries ionizing effect of a particle interaction};
\item {\bfseries scintillation light from electronic excitation};
\item {\bfseries thermal signal from lattice vibrations}.
\end{itemize} 

A great number of experiments performing one or a combination of the
mentioned techniques have been realised throughout the years, mainly
divided into two categories: single and double modality. Concerning
the former, such experiments must work in ultra-low background
conditions because they are not able to perform background
rejection. Each event, which falls in the acceptance window of the
detector is accepted as a DM candidate with its own deposit energy
measurement.  For instance CREEST I and CoGeNT belongs to this class of
experiments. CRESST I was the first phase of the CRESST experiment and
consisted of a cryogenic bolometer based on saphire crystals as target
material operating below 10 mK. This technique provides a threshold
energy on nuclear recoils of the order of 0.5-0.6 keV, much lower than
the minimum thresholds set by other experimental methods. Anyway the
reduced target mass (262 g) limit its sensitivity. The CoGeNT
experiment uses a single, 440g, high-purity germanium crystal
cooled to liquid nitrogen temperatures in its measurements. The
detector has the advantage of a very low energy threshold ($<$0.5 keV)
which allows it to search for nuclear recoil events due to DM
particles of relatively low mass ($>$ $GeV/c^2$). In addition to a
low-background configuration, the detector is able to distinguish and
reject background events from the surface by measuring the risetime of
the detector's signals. The CoGeNT detector is sensitive only to the
ionisation charge from nuclear recoils and sets limits on the mass and
interaction cross-section of DM particles by excluding any candidate
mass and cross-section pair that would result in a signal above the
background of the detector.

The CoGeNT collaboration has recently announced their results on 15 months of data, including the
measurement of the spectrum of nuclear recoil candidate events, and
the time variation of those events  \citep{CoGeNT-3-1106.1066v1}.
 These results appear consistent
with the signal anticipated from a relatively light DM particle
scattering elastically with nuclei. The observed spectrum and rate is
consistent with originating from DM particles with a mass in the range
of 4.5-12 GeV and an elastic scattering cross section with nucleons of
approximately $\sim 10^{-40} cm^2$ \citep{CoGeNT201111061066v1}.  In
early 2010, the CoGeNT collaboration reported the observation of
$\sim$100 events above expected backgrounds over a period of 56 days,
with ionisation energies in the range of approximately 0.4 to 1.0 keV
\citep{CoGeNT-1-1106.1066v1}.  Concerning the double modality, all
the experiments able perform energy measurements and background
discrimination of the electron recoils, belong to this category.  The
most important results come from CDMS and EDELWEISS. The Cryogenic DM
Search (CDMS in the Soudan Underground Laboratory) has developed a
ionisation/phonon technique which allows a high efficient event by
event discrimination between electron and nuclear recoils, based on
the simultaneous record of inciting and phonon signal in the Ge and
Si detectors packed in towers, operating at 40 mK. The recoil energy
threshold is 10 keV. Recently, results have been presented from a
re-analysis of the entire five-tower data set acquired with an
exposure of 969 kg-days \citep{e112002} with a recoil energy extended
to 150 keV and an increased sensitivity of the experiment to the
inelastic DM (iDM) model. Three DM candidates were found between 25
keV and 150 keV where the probability to observe three or more
background events in this energy range is 11$\%$. The CDMS program
goes on with a future installation of the SuperCDMS setup at the new
deep SNOLab laboratory, with 100 kg of detectors. EDELWEISS experiment
(in the Modane Underground Laboratory) is conceptually based on the
same idea as CDMS, with 300g germanium monocrystals as target. The
typical recoil energy ranges from a few keV to few tens of keV. This
recoil is measured at the same time with a heat channel, and a
ionisation channel. The energy threshold reaches 10 keV. The
EDELWEISS-II experiment has carried out a direct WIMP search with an
array of ten 400 $g$ Inter-Digit detectors, achieving an effective
exposure of 322 $kg\,days$. 
 The best sensitivity achieved in the elastic
spin-independent WIMP-nucleon cross-section is $5 \times 10^{?8} pb$ for a WIMP mass of 80 GeV .  
The results are detailed
in \citep{EDELWEISS}.

Exploiting the noble gas properties at low temperatures, both
scintillation and ionisation signal can be detected by noble liquid
detectors. Argon and Xenon at below 88K and 165K respectively behave
as dense liquids with good scintillation yield of about 40$\times
10^3$ photons/MeV and good electron mobility. The particle interaction
in the liquids ionises the medium and produces excited states of the
gas atoms, which generate a luminescence signal. There are two excited
states both for Ar and Xe (a singlet and a triplet) which differ in
lifetime, quite enough to be measured by using a pulse shape
discrimination analysis. The noble liquid detectors operate in a dual
phase mode: the noble element (in the form of liquid and gas state) is
saved in a vessel equipped with an array of photomultiplier. The
interaction within the liquid phase produces a first direct
scintillation signal, called S1, while, the produced ionisation
electrons are drifted toward the liquid-gas interface, by means of a
strong electric field applied in the volume. A second light signal S2
is emitted due to the difference of amplitude of the electric fields
in both phases. By measuring the relative timings of the different
signals, the position of the primary interaction may be
reconstructed. Furthermore, since the ionisation yield is smaller for
nuclear recoils, the S2/S1 ratio is used to distinguish nuclear from
electron recoils.  

The XENON10 experiment uses this technology, with PMT arrays both in
the gas and liquid phase. The nuclear recoil energy threshold is 4.5
keV and the effective event rate in region of interest (4.5-29.6 keV)
after all fiducial cuts is around $2 \times 10^{-3}$ /keV/kg/day
\citep{Xenon10-2008}.

Xenon10 Collaboration has reported results of a search for light DM
particle ($\lesssim$ 10 GeV) with a sensitivity threshold of 1.4
keV. Considering spin-independent dark matter- nucleon scattering, a
cross sections $\sigma_n > 3.5 \times10^{-42} cm^2$ is excluded, for a
dark matter particle mass $m_{\chi}$= 8 GeV \citep{Xenon10-2011}.

XENON100 is one of the successor projects of XENON10. The experimental
setup has been enlarged and much care has been taken to reduce the
background. The XENON100 experiment has recently completed a DM run
with 100.9 live-days of data, taken from January to June 2010.
%Events in a
%48 kg fiducial volume in the energy range between 8.4 and 44.6
%keV$_{nr}$ have been analyzed. 
A total of three events have been found
in the fiducial volume analyzed, compatible with the background
prediction of (1.8 $\pm$ 0.6) events \citep{1104.3121v1}.

The WARP collaboration has built a dual-phase prototype liquid Argon
with a fiducial mass of 1.8 kg which achieved a good discrimination of
the $^{39}$Ar background and reached a background rate of $2 \times
10^{-3}$ /keV/kg/day after discrimination in the 20-40 keVee
\footnote{keV electron-equivalent.} range. The WARP
collaboration has also produced a large (order of 100 kg) detector at
Gran Sasso with a massive active liquid argon shield, but unforeseen
technical difficulties seem to prohibit a timely start of the
experiment.  In the dual phase detectors the major part
of the discrimination power comes from the pulse shape. On this basis,
a single phase liquid Argon project has been proposed
and two detectors, DEAP 3600 and MiniClean with fiducial masses of
1000 kg and 100 kg respectively are under construction at SNOLAB. The
pulse-shape based discrimination is expected to provide a sufficient
electron recoil background reduction to completely suppress the
intrinsic radioactivity, but it is also being considered to fill the
detector with argon extracted from underground sources which are
depleted in $^{39}$Ar by a factor of 20 or more.  

In the last years a new experiment, called DarkSide \citep{DarkSide} 
has been proposed by a Chinese, Italian, Russian and US Collaboration. It consists of a large
depleted argon TPCs for direct DM searches at LNGS.
The innovations proposed make DarkSide a detector of unprecedent 
background-free performance. The mechanical design is in
progress. The installation will start in summer 2012. 
The detector will reach, in 3 years background free operation, a cross-section sensitivity of 
$1.5 \times 10^{-45}$ cm$^{2}$ for WIMP-nucleon scattering, competitive with
the projected sensitivity of other present experiments.

Other projects are based on superheated liquids, used in particle
detectors early on in the form of bubble chambers. The PICASSO project
is based on this technology, but instead of a monolithic bubble
chamber, the detectors consist of tiny droplets immersed in a gel
matrix. In appropriate temperature and pressure conditions, the
interacting particle creates a bubble in the target volume. The
operation conditions can be tuned in a way that the detector is
sensitive only for nuclear recoils while is essentially blind to gamma
rays and cosmic muon induced events with small energy deposit.  The
advantage of this technology are the relatively low costs and simple
detector production, while a clear disadvantage is that there is no
energy information available. PICASSO has operated several of these
detectors with a total volume of 4.5 l, containing of order of 70 g of
the main WIMP target fluorine each. With this target the experiment is
mainly sensitive to spin-dependent WIMP-nucleon interactions. PICASSO reported results on the limits obtained on the
spin-dependent cross section for WIMP scattering on $^{19}F$, setting
a limit for WIMP interactions on protons of 0.16 pb (90$\%$ CL).

The largest operating experiment for DM direct search is DAMA \citep{berna2004} which is
based on about 10 kg of scintillating radiopure NaI(Tl) crystals
readout by photomultiplier's in a well shielded and controlled
environment. Although the scintillation light is detected, a signal
shape discrimination can be performed at a reasonable level so the energy
evaluation can be coupled to a discrimination method. The large mass
and low background allow to investigate the presence of DM exploiting
the model independent annual modulation signature.

%{\bf DAMA}

%The largest operating experiment for direct DM search is
%DAMA. 

%The DAMA project is based on the development and use of low background
%scintillators, and several low background set-ups have been realized
%and used for various kinds of investigations [1]. In particular, the
%former DAMA/NaI and the present DAMA/LIBRA experiments at the Gran
%Sasso National Laboratory of the INFN have the main aim to investigate
%the presence of DM (DM) particles in the galactic halo by
%exploiting the model independent DM annual modulation signature.

%The DAMA/LIBRA target consists of large ($\sim$ 10 kg) scintillating
%NaI(Tl) crystals read out by photo-multiplier tubes (PMTs) at room
%temperature in a well shielded and controlled environment. 
Since NaI does not provide a strong event-by-event discrimination of
electron recoil background, DAMA follows a unique strategy to still
get a handle on the background: the Sun orbits our galaxy with a
velocity of $\sim$ 220 km/s, while the earth rotates around the sun
with $\sim$ 30 km/s. Assuming the DM halo around our galaxy has no net
angular momentum, the relative velocity of WIMPs respect to the
detector changes over the course of the year. For a given energy
threshold this would lead to an annual modulation of the interaction
rate with a known phase. DM particles from the Galactic halo are hence
expected to show an annual modulation of the event rate induced by the
Earth's motion around the Sun \citep{modulation}.  DAMA reported an
effect of annual modulation of the count rate at low energies, which
attributes to a WIMP signal at 6.3$\sigma$. This
annual fluctuation in the background rates occurs near threshold, in
the 2-6 keV region where the pulse shape discrimination start to fail. The
variation is well fitted by the cosine function expected for the WIMP
signal with a period $T=0.999 \pm 0.002$ year, i.e. very close to one
year, with a phase $t_0=146 \pm 7$ days, which is very close to the
signal modulation expected for WIMPs (152.5 days or 2nd of June), and
with an amplitude of 0.0131 ev/kg/keV/day. The observed modulation
amplitude is (0.0200 $\pm$ 0.0032) cev/kg/keV/day with a phase of t$_0$ = (140$\pm$22) days and
a period of (1.00$\pm$0.01) year. In the meanwhile, the DAMA
collaboration has upgraded the detector to 250 kg NaI(Tl) called
DAMA/LIBRA which started the operations in 2003. The combination of
all DAMA/NaI and DAMA/LIBRA data shows a statistically compelling
(8$\sigma$) modulation signal at low energies (2-4 keV), compatible
with a WIMP interpretation. The DAMA group has not found any
modulation effect in the region of energies above 90 keV. If the
effect is caused by variation of the Compton background in the $2\div6$ keV
energy bin it must reveal itself also for the gammas with energies
$>$90 keV.  However, other experiments have excluded the region of the
allowed parameters $M_W$ and $\sigma_p$ derived by DAMA from the
modulation effect.

\begin{figure}[!htb]
\begin{center}
\includegraphics[width=80mm]{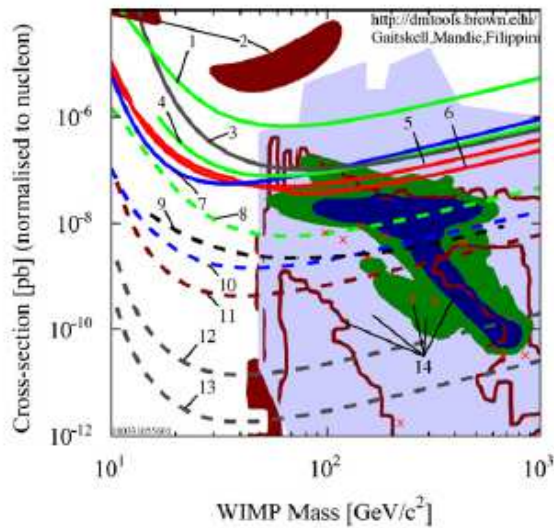}
\caption{{Experimental results and SUSY predictions for WIMP nucleon cross sections 
versus WIMP mass 1-ZEPLIN-II, 2-DAMA/LIBRA,
  3-EDELWEISS, 4-ZEPLIN-III 1st phase, 5-CDMS,
  6-CDMS, 7-Xenon10, 8-ZEPLIN-III 2D phase, 9-XMASS
  , 10-Xenon100 , 11-LUX, 12-Xenon1t , LZS ,
  MAXG2LXe, 13-LZD, MAXG3LXe and 14-SUSY predictions.}}
\label{plotexcl}
\end{center}
\end{figure}

The experiments on WIMP search over the last years have put the limits
on the spin-independent cross-section already below 10$^{-7}$ pb. The
ultimate goal of DM search experiments is to reach
sensitivities down to 10$^{-12}$ pb. This will allow to probe the
whole $M_W$ and $\sigma_p$ parameter space of SUSY predictions. This
goal can be achieved only with the use of next-generation detectors of
the ton scale and ultimately of the multi-ton scale. 

%%%%%%%%%%%%%%%%%%%%%%%%
\subsection{Indirect Detection}
%%%%%%%%%%%%%%%%%%%%%%%%%%%%%%%

After freeze out, DM pair annihilation becomes largely
 suppressed. However, even if its impact on the DM relic
 density is negligible, DM annihilation continues and may be
 observable. Thus, DM particle are expected to either annihilate or
decay into SM particles like $\gamma$-rays, neutrinos and
matter-antimatter particles in the final state
% (see Fig. \ref{fig:DMdecays}) 
 and the detection of annihilation products is
usually referred to as indirect DM detection.

%\begin{figure}[!htb]
%\begin{center}
%\includegraphics[width=80mm]{./DMdecays.eps}
%\caption{Standard Model particles production scheme for DM
%annihilation. Reproduced from Ref.[18]. If the DM particle
%self-annihilate and produce quarks, leptons and gauge bosons, then an
%indirect search is possible by searching for the secondary
%products.({\bf estratta da FermiLAT})}
%\label{fig:DMdecays}
%\end{center}
%\end{figure}

Indirect searches look for the excesses of annihilation products in
the diffuse background or from point sources. In particular,
concerning charged products, the search is devoted to the antimatter
component due to the large amount of primary matter component in the
cosmic rays.

There are several methods for DM indirect detection. Their
 relative sensitivities are highly dependent on what WIMP candidate
 is being considered, and the systematic uncertainties and
 difficulties in determining backgrounds also vary greatly from one
 method to another.

High WIMP density regions, e.g. the centers of galaxies, but also the
core of astrophysical objects such as Sun and Earth, are expected to
accumulated large amount of WIMPs. When WIMPs pass through the Sun or
the Earth or the galactic center, they may scatter and be slowed below
escape velocity. In this way, they may become gravitationally bound in
these gravitational wells. Over the
age of the solar system, their densities and annihilation rates are
greatly enhanced.

Although $\gamma$-rays, positrons and anti-protons produced in these
annihilations do not escape the Sun or the Earth, neutrinos would be
able in this because of their high penetration ability.

These neutrinos can travel to the surface of the Earth, where they may
convert to charged leptons through $\nu q \rightarrow l q'$ and the
charged leptons may be detected. The resulting neutrino flux could be
detectable as a localised emission with earth-based neutrino
telescopes by exploiting the Cherenkov light produced by the charged
lepton.

Searches for neutrinos are unique among indirect searches since they
are, under certain assumptions, probes of scattering cross sections,
not annihilation cross sections, and so are competitive with the
direct detection searches.

In the Sun both spin-independent as well as spin-dependent scattering
can lead to the capture of DM. Among nuclei with a net spin,
hydrogen is the only one present in the Sun in significant
proportions. Other trace elements in the Sun generally have no net
spin, and even when they do, there is no $A^2$ enhancement in the
cross section coherent term because of their low density
fraction. Therefore, the only relevant quantity for spin-dependent
capture is the WIMP-proton cross section. Spin-independent capture, on
the other hand, receives contributions from several elements. In fact,
the cross section for spin-independent WIMP-nucleus scattering is
strongly enhanced by large A. Therefore, even though the heavier
elements in the Sun are rarer, this enhancement makes their
contribution to the capture rate significant. In particular, oxygen
plays the most important role in spin-independent capture of WIMPs in
the Sun.

Neutrino telescopes could probe direct search
observation. In fact, the signals observed at the direct detection
experiments DAMA, CoGeNT and CRESST could be explained by light WIMPs
with sizeable spin-independent cross sections with nucleons. The
capture and subsequent annihilation of such particles in the Sun would
induce neutrino signals in the GeV range which may be observed at
Super-Kamiokande\citep{superk}.

Actually, detection of these neutrinos in the range $10 GeV < E_\nu <
1 TeV$ by large neutrino detectors such as SuperKamiokande ,
AMANDA  \citep{AMANDA}  and IceCube \citep{ICECUBE} so far provides only upper limits on the high
energy neutrino flux from the center of the Sun or the Earth (see \citep{kundu}).

Future neutrino searches at Super-Kamiokande may
have lower thresholds and so provide leading bounds on low mass WIMPs.
In this way, Super-Kamiokande may test the DAMA and CoGeNT signal
regions at high $\sigma_{SI}$ and $m_{X} \sim$ 1 - 10 GeV
\citep{Hooper, Feng, Kumar}. 

Moreover installing DeepCore, also IceCube significantly lowers its
energy threshold and enhances the ability of detecting neutrinos from
light WIMP annihilation.

Unlike other indirect DM searches this method does not depend
strongly on our galaxy's DM halo profile or on the
distribution of DM substructure.

The event rate depends on the DM density in the Sun, which in
turn is dictated by the cross section of WIMPs with nucleons. This is
constrained by direct detection experiments. The neutrino flux depends
on the WIMP density, which is determined by the competing processes of
capture, that is from the scattering cross section, and annihilation.
Moreover the differential neutrino flux depends also on the way in
which neutrinos are produced. Assuming that WIMPs annihilate to $b
\bar{b}$ or $W^+ W^-$ which decay to neutrinos, as in many neutralino
models, the neutrino signal is completely determined by the scattering
cross section. For the majority of particle physics models considered
(e.g., supersymmetry or KK models), the WIMP capture and annihilation
rates reach or nearly reach equilibrium in the Sun. This is often not
the case for the Earth. First, the Earth is less massive than the Sun
and, therefore, provides fewer targets for WIMP scattering and a less
deep gravitational well for capture. Secondly, in the Earth
spin-independent interactions may occur. For these reasons, it is
unlikely that the Earth will provide any observable neutrino signals
from WIMP annihilations in any planned experiments.

If the annihilation occurs in free space, other types of radiation can
be detected.
The DM particles annihilating within the halo, can arise production of
SM particles associated with the emission of $\gamma$ ray with
energies of the order of the DM particle mass. Thus, the $\gamma$-ray
emission associated with such annihilation provides a chance for DM
particles detection.
The most striking gamma ray signal would be mono-energetic
 photons from $\chi \chi \rightarrow \gamma \gamma$, but since WIMPs
 cannot be charged, these processes are typically loop-induced or
 otherwise highly suppressed. More commonly, gamma rays are produced
 when WIMPs annihilate to other particles, which then radiate
 photons, leading to a smooth distribution of gamma ray energies. If
 gammas arise together with other particles, we only would expect
 some enhancement of the spectrum below the WIMP mass. The main
 difficulty in the extraction of information about DM from
 the annihilation photons is the presence of large and uncertain
 gamma-ray backgrounds. On the other hand, photons point back to
 their source, providing a powerful diagnostic. Possible targets for
 gamma ray searches are the center of the Galaxy, where signal rates
 are high but backgrounds are also high and potentially hard to
 estimate, and dwarf galaxies, where signal rates are lower, but
 backgrounds are also expected to be low.
Moreover, for most particle DM candidates, $\gamma$-rays
would be accompanied by neutrinos, thus neutrino experiments could
help to confirm whether the $\gamma$ flux originates from DM
annihilations or from other astrophysical sources.

Another difficulty for DM search with $\gamma$ arises from the fact
that the DM gamma ray emission is expected to be a function of the DM
density profile, which is not experimentally known. The most adopted
density profile models are the Navarro-Frenk-White profile for which:

\begin{equation}
\rho(r) = \frac{\rho_0}{\frac{r}{R_s}(1+\frac{r}{R_s})^2}
\end{equation}

where $\rho_0$ and $R_s$, are parameters which vary from halo to halo,
and the Einasto profile for which:

\begin{equation}
\rho(r) \propto exp (-A r^\alpha)
\end{equation}

The parameter $\alpha$ controls the degree of curvature of the
profile.

The expected energy integrated $\gamma$-ray flux from a region of
volume V at a distance D with DM density $\rho$ is:

\begin{equation}
\phi_{\gamma} = N_{\gamma} \frac{\langle \sigma v \rangle}{m^2} \frac{1}{4 \pi D^2} \int_V dV \frac{\rho^2}{2}, 
\end{equation}

where $N_\gamma$ is the number of $\gamma$-rays per collision, $m$ is
the DM particle mass and $\langle \sigma v \rangle$ is the
velocity-weighted annihilation cross-section. The current cosmological
DM density is set by their annihilation rate in the early
Universe. This provides a natural value for the annihilation
cross-section of $\langle \sigma v \rangle \sim 3 \times 10^{-26} cm^3
s^{-1}$.

Obviously, the higher $\rho$ the better, so that the annihilation
radiation is searched for in the most dense regions of the Milky Way
(MW).

The improvement of the Galactic diffuse model as well as the potential
contribution from other astrophysical sources could provide a better
description of the data.

$\gamma$-ray observations, in particular, at very high energies where
DM masses of a few 100 GeV and above are probed, are a promising way
to detect DM by space-based experiments, such as Fermi and
AMS, and by ground-based atmospheric Cherenkov telescopes.

H.E.S.S. is an array of four Imaging Atmospheric Cherenkov Telecopes
(IACTs) that uses the atmosphere as a calorimeter and images
electromagnetic showers induced by TeV $\gamma$-rays \citep{hof}.

H.E.S.S. Collaboration performed searches for a very-high-energy (VHE;
$\ge$ 100 GeV) $\gamma$-ray signal from annihilation of DM particles,
towards the Galactic center, the Sagittarius Dwarf galaxy and
hypothetical DM spikes that could have formed around Intermediate Mass
Black Holes (IMBHs).

The $\gamma$-ray energy spectrum in the range between 300 GeV and 30
TeV after the background suppression shows no any convincing evidence
for $\gamma$-ray emission from self-annihilating DM. Limits on
annihilating cross section for $\chi$ and $B_1$ have been found
constraining some models.

For instance, limits on $\langle \sigma v \rangle$ are derived as a
function of the DM particle mass assuming Navarro-Frenk-White and
Einasto density profiles. In particular, for the DM particle mass of
$\sim$ 1 TeV, values for $\langle \sigma v \rangle$ above 3 $\times
10^{-25} cm^3 s^{-1}$ are excluded for the Einasto density profile
\citep{Abra}.

The Galactic center (GC) region is the main target for DM search,
because of its proximity and its expected large DM concentration.
However, the search for DM induced $\gamma$ rays in the GC is affected
by a strong astrophysical background. In fact, in the GC there is the
compact $\gamma$-ray source HESS J1745-290 \citep{Abra}, coincident
with the position of the supermassive black hole Sgr A* and a nearby
pulsar wind nebula, which represents other sources of VHE
$\gamma$-rays in the observed region, thus complicating the detection
of $\gamma$-ray emission from DM annihilation.

VERITAS is a ground based Cherenkov telescope for gamma ray
astronomy and has among other things searched for an enhancement of
gamma rays from the center of neighboring dwarf galaxies without a
positive signal so far \citep{VERITAS}.

Data from the EGRET satellite in the energy range 20 $MeV$ - 30 GeV,
interpreted as evidence for DM \citep{deboer}, are not confirmed by
recent data from Fermi satellite \citep{abdo}.

The new precise data from the FERMI satellite on the diffuse gamma ray
data are well-described at intermediate latitudes by standard CR
physics assuming isotropic propagation models , but towards the
Galactic center a 20\%-30\% excess has been observed in the 2-4 GeV
range or even more significantly, within $\approx 1^\circ$ from the
Galactic Center (GC), that could be interpreted as a signal of DM
annihilation (DMA) for a very light neutralino of about 8 GeV
\citep{deboer11}.

%%%%%%%%%%%%%%%%%%%%%%%%%%%%%%%%%%%%%%%%%%%%%%%%%%%%%%%%%%%%%%%%%%%%%%%

%%%%%%%%%%%%%%%%%%%%%%% Antimatter %%%%%%%%%%%%%%%%%%%%%%%%%%%%%%%%%%%%%%%%%
Other products of DM annihilation are pairs of
particles-antiparticles, like $e^+e^-$, $p\bar{p}$ and $d\bar{d}$
pairs.

Since anti-matter is a very rare product of conventional sources of
cosmic radiation, is suitable as possible indicator for DM
annihilation.

Actually, unlike neutrinos and $\gamma$, positrons and anti-protons do
not point back to their sources making an unambiguous separation from
backgrounds very difficult.

In contrast to direct detection, many anomalies have been reported in
the indirect search with anti-matter, and some of these have been
interpreted as possible evidence for DM. The most prominent recent
example is the detection of positrons and electrons with energies
between 10 GeV and 1 TeV by the PAMELA, ATIC, and Fermi LAT
Collaborations \citep{PAMELApositron, ATICwebsite, FERMILATwebsite}.

PAMELA \citep{PAMELA}, a satellite based instrument specifically
designed to search for anti-particles in cosmic radiation, has
reported an enhancement of positrons around 100 GeV over the standard
expectation from diffuse galactic cosmic ray secondary models, in
which positrons result from inelastic collisions of primary protons on
the intergalactic gas nuclei.

Since DM annihilation produces as many positrons as electrons, with
particles of energies close to the DM particle mass (in the
range 10 GeV - 100 GeV for best motivated particle models), the
positron fraction is enhanced.

If such enhancement originates from DM annihilation, also a respective
enhancement of anti-protons would be expected, but is not
observed. Special DM models could
avoid hadron production, but also astrophysical explanations exist as
well, such as nearby pulsars.

Additionally, the ATIC balloon-borne experiment also reported an
anomalous ``bump'' in the total flux of $e^+ e^-$, at energies of
$\sim$ 600 GeV \citep{panov} which also has been interpreted as possible
evidence for DM annihilation. It would imply a large
mass and a very large pair annihilation cross section for DM.

Recently the experimental information available on the Cosmic Ray
Electron (CRE) spectrum has been dramatically expanded as the
Fermi-LAT Collaboration has reported a high precision
measurement of the electron spectrum from 7 GeV to 1 TeV performed
with its Large Area Telescope (LAT) \citep{FermiLAT}.

The Fermi experiment, like also ATIC, is unable to distinguish
positrons from electrons, and so constrain the total $e^+ + e^-$ flux.

The ATIC ``bump" is not confirmed by the Fermi LAT data.
The Fermi-LAT spectrum shows no prominent spectral features and the
spectral index of the cosmic ray background is significantly harder
than that inferred from several previous experiments: depending on the
diffusion model, best fit injection spectral indexes range between 2.3
and 2.4, as opposed to previous models with 2.54. These data together
with the PAMELA data on the rise above 10 GeV of the positron fraction
 are quite difficult to explain with just secondary production
. An additional primary positron source is required to match the
high-energy positron fraction, but other astrophysical explanations
 could be considered.

Additional data from, for example, Fermi and the Alpha Magnetic
Spectrometer (AMS) \citep{AMS02}, an anti-matter detector placed on the
International Space Station, may be able to distinguish the various
proposed explanations for the positron excesses, as well as be
sensitive to canonical WIMP models, but it remains to be seen whether
the astrophysical backgrounds may be sufficiently well understood for
these experiments to realize their DM search potential.

Other promising indirect detection search strategies are based on
anti-protons and anti-deuterons from WIMP annihilation in the galactic
halo. This searches are sensitive to DM candidates that annihilate
primarily to quarks, but, until now, no result is reported \citep{AMS,
  PAMELApbar}.

%%%%%%%%%%%%%%%%%%%%%%%%%%%
\subsection{Particle Colliders}
%%%%%%%%%%%%%%%%%%%%%%%%%%%%%%%%%%%%%%

Contributions to solve the mystery of the DM will come also from
colliders. Collider experiments can access interactions not probed by
direct detection searches.  A suitable room to search for DM is SUSY
search and presently LHC data and analyzes are advancing very quickly,
and possible signal for direct detection of supersymmetry are strictly
related to the DM discussion.

The LHC is a $p - p$ collider, designed to operate at a center of mass
energy up to $\sqrt s =14$ TeV \citep{lhc}.  Two main general purpose
detectors are installed at the LHC: ATLAS and CMS. They are designed
to perform precision measurements of photons, electrons, muons and
hadrons. At this writing the LHC is colliding p-p at $\sqrt s =7$ TeV
and analyzes of about 1 fb$^{-1}$ has already excluded squarks with
mass up to 1 TeV about, in many models, superseding results from
Fermilab-Tevatron $p \bar p$ collider with $\sqrt s =1.96$ TeV (CDF
\citep{cdf} and D0 \citep{d0} experiments). At hadron colliders heavy
particles can be produced via quark-antiquark collisions, as at
Tevatron, or via gluon-gluon and quark-gluon collision as at the LHC:
the signals are an inclusive combination of all these processes.  At
Tevatron valence quark annihilation into virtual weak bosons give up
to chargino and neutralino production with a large cross-section,
unless squarks or gluino were lighter then 300 GeV, value ruled out by
LHC.  At LHC production of gluinos and squarks by gluon-gluon and
gluon-quark fusion should be dominant.  At hadron colliders,
independently of whether it is a $p p$ or $p \bar p$ kind, one can
also have associated production of a chargino or neutralino together
with a sqaurk or gluino, but most models predict diverse cross-section
for several processes: slepton pair production was quite small at
Tevatron, but might be observable eventually at the LHC.

Produced sparticles decay in final states with two neutralino LSPs,
which escape the detectors, carrying away a part of the missing energy
which is at least two times the mass of neutralinos.  At hadron
colliders only the component of the missing energy that is manifest as
momenta transverse to the colliding beams is observable, so generally
the observable signals for supersymmetry are: $N$ leptons and $M$ jets
with a missing transverse energy, where $N$ or $M$ might be zero.

The main problem is that significant backgrounds to these signal come
from Standard Model particle, in particular processes involving
production of W and Z which decays into neutrinos, providing missing
transverse energy.

If LHC discovers signatures which seems like R-parity conserving SUSY
signatures, a SUSY mass scale has to be derived from the deviation
from SM. Early LHC data had excluded a SUSY mass scale in the sub-TeV
range \citep{PMartin}. SUSY contribution to DM from excess above
the SM, observed in observables, should be characterized by the decay
chains isolated to disentangle this contribution.  The amount of LSP
relic density of the Universe today is determined by the efficiency of
LSP annihilation before freeze-out, that is by the cross section for
$\chi \chi \to f \bar{f} / W^{+} W^{-} / ZZ$ \citep{diciaccio}.
ATLAS/CMS data can determine the LSP relic density, once one have
estimated the SUSY parameters which enter the dominating annihilation
process/processes. In mSUGRA model \citep{msugra} the LSP is usually
most bino and in this case the basic annihilation process $\chi \chi
\to f \bar{f}$ proceed via exchange of a slepton or squark and has a
too low cross-section.  To determine the LSP composition, the mass of
additional neutralinos are important, being SUSY models dependent from
many parameters as mass of LSP, sleptons/quarks and ,whether they are
superpartners of left-handed or right-handed SM particles, the neutral
Higgs masses.  Up to now no deviation from SM has been observed

%%%%%%%%%%%%%%%%%%%%%%%%%%%%%%
\section{Discussion of Experimental Results}
As we have shortly reported, during the last years, several experiments  have performed direct and indirect search investigating the possibility of DM presence.
  
Concerning the direct DM detection, 
recently, the CRESST-II and CoGeNT collaborations have reported the
observation of low-energy events in excess over known backgrounds. This result has encouraged the hypothesis that such signals in
addition to the long standing DAMA annual modulation signal 
might arise from the scattering of a light ( $\lesssim$10 GeV) DM particle.

The rate of DM elastic
scattering events   as a result of  Earth's motion around the Sun and relative to the
rest frame of the DM halo, is predicted to vary with an annual cycle.
The only experiment reporting the observations of an annual modulation is DAMA/LIBRA which detected with
high significance  
($\sim 8 \sigma$)  with a phase and period
consistent with elastically scattering DM. 

The spectrum
of the signals reported by DAMA/LIBRA and CoGeNT seem to point toward a
similar range of DM parameter space. Moreover, the
range of DM mass implied by CoGeNT and DAMA/LIBRA is very
similar to that required to explain the spectrum of gamma rays
observed by the Fermi Gamma Ray Space Telescope from the the inner
$0.5^\circ$ around the Galactic Center \citep{8}, and for the observed
synchrotron emission known as the WMAP Haze \citep{9}.
Anyway much care must be taken for this interpretation since CoGeNT experiment does not discriminate between electron and nucleon recoils \citep{Ccoll}.

Against to the positive signals of DAMA and CoGeNT,
several other experiments find no evidence for DM. In particular, the
CDMS, Xenon10 and Xenon100 collaborations
disfavor the parameter space indicated by DAMA and CoGeNT.

One remarkable difference between DAMA/CoGeNT and these
experiments, is that the latter reject electronic interactions
attempting to collect only nuclear recoil events. In fact, it would be possible 
that the anomalous signals arise from such electronic recoils, a
possibility that would explain away the existing tension. 
A model of
this type was considered in prior to the recent CoGeNT
measurement and it remains to be seen whether this possibility is
theoretically feasible.
Concerning the inelastic DM interpretation the spin independent fit to the DAMA modulated rate allows two
qualitatively different best-fit regions: one around $M_{DM} \approx
80 GeV$ with $\sigma \approx 10^{-41} cm^2$ due to scattering on
iodine (A = 127, Z = 53) and one around $M_{DM} \approx 10 GeV$ with
$\sigma \approx 10^{-40} cm^2$ due to scattering on sodium (A = 23, Z
= 11). The first region is firmly excluded by many other experiments,
such XENON100 and CDMS. The second region, while still disfavoured by
other null searches, is not completely excluded due to many experimental
uncertainties and due to the general difficulty of direct detection
searches to deal with low recoil energy scatterings.
A possibility to make the DAMA results consistent with other experiments is to include an effect called "channelling" which will be present only the NaI crystals with DAMA uses. Anyway though inclusion of this effect the situation does not improve significantly. For a WIMP mass close to $10GeV$  with the help of NaI channelling it is possible to explain the DAMA results in terms of spin independent inelastic DM nucleon scattering.
In that case some relevant parameters as DM mass and splitting should be fine tuned, and also the WIMP velocity distribution in the Galaxy should be close to the escape velocity.
Inelastic spin-dependent interpretation would be a possibility which do not receive significant constraints from other experiments. Anyway both interpretations make difficult to reconciling DAMA and CoGeNT results \citep{Ccoll}.

A possibility to avoid the conflict between DAMA results and other  experiments is proposed in terms of different but not irrelevant DM candidates. Stable particles with charge -2, bound with primordial helium would represent a kind of nuclear interacting form of DM. Such particles cannot be detected using nuclear recoils by direct search experiments  , but their low energy binding with Na nuclei can  the annual variations of energy release from their radiative capture, in the energy range 2-4 KeV corresponding to the signal observed in DAMA and DAMA/LIBRA \citep{khlo1,khlo2,khlo3}.

Regarding indirect detection
 four experiments on DM search have generated great excitement in the astroparticle community. Their results are briefly summarized below:

\begin{itemize}
\item the PAMELA experiment measured with high statistics the positron
  component in the cosmic ray flux in the energy range between 10 and
  100 GeV, observing a positrons excess above 10 GeV claimed also by
  earlier experiments as HEAT and AMS whose experimental data extend
  up to 40 GeV but with larger uncertainties. Anyway the rise up to 20
  GeV can be explained by solar modulation which depends on the charge
  sign of a particle affecting positrons and electrons in a different
  way \citep{PAMELApositron}. No excess in the antiproton/proton
  fraction has been observed by PAMELA \citep{PAMELApbar}.  The positron excess cannot be
  explained by a purely secondary production (due to primary protons
  and He nuclei interacting with the interstellar medium), which is
  characterized by a not so hard spectrum suggesting the existence of
  other primary sources.

\item The ATIC ballon born experiment has measured the total $e^+ + e^-$
  spectrum in the energy range 3 GeV-2.5 TeV, finding an increase of
  the flux from 100 GeV up to 600 GeV where they observed a peak
  followed by a sharp fall to about 800 GeV. The rest of the data
  agree with the GALPROP theoretical predictions within the errors.

\item The FERMI experiment is able to measure with high resolution and
  statistics $\gamma$-rays in the energy range 20 MeV-300 GeV and
  primary cosmic ray $e^+ + e^-$ spectrum between 20 GeV and 1
  TeV. Recent results on a data sample of 6 months confirmed the
  excess in the total $e^+ + e^-$ flux seen by ATIC, even though with
  a flatter trend.

\item The large array telescope HESS measured $\gamma$-rays up to 5
  TeV. HESS's results do not confirm ATIC peak as well as the sharp
  fall around 800 GeV, but a suitable HESS data normalization leaves a
  room for an agreement with ATIC results within the uncertainties.

\end{itemize}

Several interpretations came out invoking different sources: from
those purely astrophysical like nearby pulsars or SNR to more exotic
such as DM annihilation or decay in the halo of our Galaxy, arising
from annihilation or decay of DM particles. Presently we are not able
to say which interpretation assumes more validity. Nearby pulsars have
been proposed as accelerating mechanism of energetic particles to
explain the observed positron excess.  Primary electrons should be
accelerated in the pulsar magnetosphere, emitting gamma rays by
synchrotron radiation. In the high magnetic fields of the pulsar, such
gammas induce and electromagnetic cascade. The electron-positron pairs
are then accelerated and confined in the pulsar nebula before escaping
into the interstellar medium, so enhancing the CR electron and
positron components.  The energy spectrum of such particles is
expected to be harder than that of the positrons of secondary
production , thus the positrons originated by pulsars may dominate the
high energy end of the CR positron spectrum. Two nearby pulsars could
enhance significantly the high energy electron and positron flux
reaching the Earth: Monogem at a distance of 290 pc and Geminga at a
distance of 160 pc. This because electrons and positrons, during their
propagation loose energy mainly by inverse Compton scattering and
synchrotron radiation so only sources at distance less than 1 kpc can
give a significant contribute to the energy spectrum.  The
interpretation of the FERMI, PAMELA and HESS excesses in terms of DM
signature is quite suggestive but encounters some difficulties arising
from the PAMELA data which show an asymmetry between the hadronic
(antiprotons) and leptonic (positrons) component.

A first source of disappointment for the neutralino devotees is, in this case, due to the fact that, assuming the usual local DM density $\rho = 0.3 GeV/ cm^3$, the annihilation cross section needed to explain the signal is about
$\sigma v \simeq 1 \times 10^{-24} cm^3 s^{-1}$, while the value
$\sigma v \simeq 3 \times 10^{-26} cm^3 s^{-1}$, is needed in the early
universe for a thermal WIMP in order to provide the correct relic
abundance.

 In conclusion, it seems that we are on the track to find  evidences for DM  but no {\it experimentum crucis} has given final probes. A main challenge comes from cosmology since it seems that DM behaves in different ways depending on the astrophysical structures and cosmological scales. For example, DM dynamics is very different in the various self-gravitating systems as elliptical, spiral and dwarf galaxies. Besides, it seems that different DM components have to be considered  to address local systems (galaxies) and large scale structures (clusters of galaxies and super clusters of galaxies). An alternative view, could be that the problem of missing matter does not require the introduction of new ingredient but a modification or an extension of GR. The point is  that gravity  could not work in the same way at any scale. This issue will be faced in the next section.

%%%%%%%%%%%%%%%%%%%%%%%%%%%%%%%%%%%%%%%%%%%%%%%%%%%%%%%%%%%%%%%%%%%%%%%%%%%%%%%%%%%%%%%%%%%%%%%%%%%%%%%%%%%%%%%%%%
%%%%%%%%%%%%%%%%%%%%%%%%%%%%%%%%%%%%%%%%%%%%%%%%%%%%%%%%%%%%%
\section{An alternative view}

 Both cosmic speed up (DE) and DM,  instead of being related to the search of new ingredients, could be the signal of a
breakdown in our understanding of the laws of gravitation at large (infra-red) scales. From this point of view,  
one should consider the possibility that the Hilbert\,-\,Einstein
Lagrangian,  on which GR relies and linear in the Ricci scalar $R$,  should be generalized.
Following this approach, the choice of the effective theory of gravity could  be derived by means of the data and the "economic"
requirement that no exotic ingredients have to be added. This is
the underlying philosophy of what is referred to as Extended Theories of Gravity consisting in enlarging the geometric sector of GR and assuming the possibility that other curvature invariants could contribute to the dynamics
\citep{booksalv, reportnoi}. From a
theoretical standpoint, several issues from fundamental physics
(quantum field theory on curved spacetimes, M-theory etc.) suggest
that higher order terms must necessarily enter the gravity
Lagrangian. On the other side, Solar System experiments show the
validity of Einstein's theory at local  scales so the problems could come out at galactic scales and beyond. 
The simplest extension of Einstein theory is the possibility to take into account actions of the form $f(R)$ where $f$ is a function of the Ricci scalar $R$ not necessarily linear in R as in the Hilbert case. This kind of theories 
 have recently received much attention in cosmology, since
they are naturally able to give rise to accelerating expansions
(both in the late and the early Universe). However, it is possible to demonstrate that
$f(R)$ theories can also play a major role at astrophysical
scales \citep{booksalv}. In fact, modifying the gravitational Lagrangian can affect the
gravitational potential in the low energy limit,
provided that the modified potential reduces to the Newtonian one
at  the Solar System scales. In fact, a corrected gravitational
potential could offer the possibility to fit galaxy rotation
curves without the need of DM. In
addition, one could work out a formal analogy between the
corrections to the Newtonian potential and the usually adopted
DM models.  The choice of an analytic function in term
of Ricci scalar is physically corroborated by the Ostrogradski
theorem, which states that this kind of Lagrangian
is the only viable one which can be considered among the several
that can be constructed by means of curvature tensor and possibly
its covariant derivatives. The field equations of this approach can be
recast in the Einstein form, that is:
\begin{equation}\label{5}
G_{\alpha \beta} = R_{\alpha\beta}-\frac{1}{2}g_{\alpha\beta}R =
T^{curv}_{\alpha\beta}+T^{M}_{\alpha\beta}/f^\prime(R)
\end{equation}
where the prime denotes derivative with respect to $R$,
$T^{M}_{\alpha \beta}$ is the standard matter stress\,-\,energy
tensor and
\begin{eqnarray}
T^{curv}_{\alpha\beta}\,=\,\frac{1}{f'(R)}\Big\{\frac{1}{2}g_{\alpha\beta}\left[f(R)-Rf'(R)\right]
+f'(R)^{;\mu\nu}(g_{\alpha\mu}g_{\beta\nu}-g_{\alpha\beta}g_{\mu\nu})
\Big\}\,, \label{6}
\end{eqnarray}
defines a  {\it curvature stress\,-\,energy tensor}. The presence
of the terms $f^\prime(R)_{;\mu\nu}$ renders the equations of
fourth order, while, for $f(R) = R$, Eqs.(\ref{5}) reduce to the
standard second\,-\,order Einstein field equations. As it is clear
from Eq.(\ref{5}), the curvature stress\,-\,energy tensor formally
plays the role of a further source term in the field equations
which effect is the same as that of an effective fluid of purely
geometrical origin. Depending on the scales, it is such a
curvature fluid which can play the role of DM and DE. From the cosmological viewpoint, in the standard framework
of a spatially flat homogeneous and isotropic Universe, the
cosmological dynamics is determined by its energy budget through
the Friedmann equations. In particular, the cosmic acceleration is
achieved when the r.h.s. of the acceleration equation remains
positive. In physical units, we have $ \ddot{a}/{a} = - (1/6) (
\rho_{tot} + 3 p_{tot} ) \,,$ where $a$ is the cosmic scale factor, $H =
\dot{a}/a$ the Hubble parameter, the dot denotes derivative with
respect to cosmic time, and the subscript $tot$ denotes the sum of
the curvature fluid and the matter contribution to the energy
density and pressure. From the above relation, the acceleration
condition, for a dust dominated model, leads to:
%\begin{equation}
$\rho_{curv} + \rho_M + 3p_{curv} < 0 \rightarrow w_{curv} < -
\frac{\rho_{tot}}{3 \rho_{curv}}$
%\label{eq: condition}
%\end{equation}
so that a key role is played by the effective quantities\,:

\begin{equation}
\rho_{curv} = \frac{1}{f'(R)} \left \{ \frac{1}{2} \left [ f(R) -
R f'(R) \right ] - 3 H \dot{R} f''(R) \right \} \ , \label{eq:
rhocurv}
\end{equation}
\begin{equation}
w_{curv} = -1 + \frac{\ddot{R} f''(R) + \dot{R} \left [ \dot{R}
f'''(R) - H f''(R) \right ]} {\left [ f(R) - R f'(R) \right ]/2 -
3 H \dot{R} f''(R)} \ . \label{eq: wcurv}
\end{equation}
As a direct simplest choice, one may assume a power\,-\,law form
$f(R) = f_0 R^n$, with $n$ a real number, which represents a
straightforward generalization of the Einstein General Relativity
in the limit $n=1$. One can find power\,-\,law solutions for
$a(t)$ providing a satisfactory fit to the SNeIa data and a good
agreement with the estimated age of the Universe in the range
$1.366 < n < 1.376$ \citep{booksalv}. It is worth noticing, that
even an inverse approach for the choice of $f(R)$ is in order.
Cosmological equations derived from (\ref{5}) can be reduced to a
linear third order differential equation for the function
$f(R(z))$, where $z$ is the redshift. The Hubble parameter $H(z)$
inferred from the data and the relation between $z$ and $R$ can be
used to finally work out $f(R)$. In addition, one
may consider the expression for $H(z)$ in a given DE
model as the input for the above reconstruction of $f(R)$ and thus
work out a $f(R)$ theory giving rise to the same dynamics as the
input model. This suggests the intriguing possibility to
consider observationally viable DM models (such as
$\Lambda$CDM and quintessence) only as effective parameterizations
of the curvature fluid \citep{mimic}.

The successful results obtained at cosmological scales motivates
the investigation of $f(R)$ theories even at astrophysical scales.
 In the low energy limit,
higher order gravity implies a modified gravitational potential.
Now, by considering the case of a pointlike mass $m$ and solving
the vacuum field equations for a Schwarzschild\,-\,like metric, one gets from a theory $f(R)=f_0 R^n$ the modified
gravitational potential:
\begin{equation}
\Phi(r) = - \frac{G m}{r} \left [ 1 + \left ( \frac{r}{r_c} \right
)^{\beta} \right ] \label{eq: pointphi}
\end{equation}
where
\begin{equation}
\beta = \frac{12n^2 - 7n - 1 - \sqrt{36n^4 + 12n^3 - 83n^2 + 50n +
1}}{6n^2 + 4n - 2} \label{eq: bnfinal}
\end{equation}
which corrects the ordinary Newtonian potential by a power\,-\,law
term. In particular, this correction sets in on scales larger than
$r_c$ which value depends essentially on the mass of the system.
The corrected potential (\ref{eq: pointphi}) reduces to the
standard $\Phi \propto 1/r$ for $n=1$ as it can be seen from the
relation (\ref{eq: bnfinal}). The generalization of Eq.(\ref{eq:
pointphi}) to extended systems is straightforward. We simply
divide the system in infinitesimal mass elements and sum up the
potentials generated by each single element. In the continuum
limit, we replace the sum with an integral over the mass density
of the system taking care of eventual symmetries of the mass
distribution. Once the gravitational potential has been computed,
one may evaluate the rotation curve $v_c^2(r)$ and compare it with
the data. For
 the pointlike case we have\,:
\begin{equation}
v_c^2(r) = \frac{G m}{r} \left [ 1 + (1 - \beta) \left (
\frac{r}{r_c} \right )^{\beta} \right ] \ . \label{eq: vcpoint}
\end{equation}
Compared with the Newtonian result $v_c^2 = G m/r$, the corrected
rotation curve is modified by the addition of the second term in
the r.h.s. of Eq.(\ref{eq: vcpoint}). For $0 <\, \beta \,< 1$, the
corrected rotation curve is higher than the Newtonian one. Since
measurements of spiral galaxies rotation curves signals a circular
velocity higher than what is predicted on the basis of the
observed luminous mass and the Newtonian potential, the above
result suggests the possibility that our modified gravitational
potential may fill the gap between theory and observations without
the need of additional DM. It is worth noting that the
corrected rotation curve is asymptotically vanishing as in the
Newtonian case, while it is usually claimed that observed rotation
curves are flat. Actually, observations do not probe $v_c$ up to
infinity, but only show that the rotation curve is flat within the
measurement uncertainties up to the last measured point. This fact
by no way excludes the possibility that $v_c$ goes to zero at
infinity.
\begin{table}
\caption{Best fit values of the model parameters from maximizing
the joint likelihood function ${\cal{L}}(\beta, \log{r_c}, f_g)$.
We also report the value of $\Upsilon_{\star}$, the $\chi^2/dof$
for the best fit parameters (with $dof = N - 3$ and $N$ the number
of datapoints) and the root mean square $\sigma_{rms}$ of the fit
residuals. See also Fig.3 \citep{cardone}.}
\begin{center}
\begin{tabular}{|c|c|c|c|c|c|c|}
\hline Id & $\beta$ & $\log{r_c}$ & $f_g$ & $\Upsilon_{\star}$ &
$\chi^2/dof$ & $\sigma_{rms}$ \\
\hline UGC 1230 & 0.608 & -0.24 & 0.26 & 7.78 & 3.24/8 & 0.54 \\
UGC 1281 & 0.485 & -2.46 & 0.57 & 0.88 & 3.98/21 & 0.41 \\ UGC
3137 & 0.572 & -1.97 & 0.77 & 5.54 & 49.4/26 & 1.31 \\ UGC 3371 &
0.588 & -1.74 & 0.49 & 2.44 & 0.97/15 & 0.23 \\ UGC 4173 & 0.532 &
-0.17 & 0.49 & 5.01 & 0.07/10 & 0.07 \\ UGC 4325 & 0.588 & -3.04 &
0.75 & 0.37 & 0.20/13 & 0.11 \\ NGC 2366 & 0.532 & 0.99 & 0.32 &
6.67 & 30.6/25 & 1.04 \\ IC 2233 & 0.807 & -1.68 & 0.62 & 1.38 &
16.29/22 & 0.81\\ NGC 3274 & 0.519 & -2.65 & 0.72 & 1.12 &
19.62/20 & 0.92 \\ NGC 4395 & 0.578 & 0.35 & 0.17 & 6.17 &
34.81/52 & 0.80 \\ NGC 4455 & 0.775 & -2.04 & 0.88 & 0.29 &
3.71/17 & 0.43 \\ NGC 5023 & 0.714 & -2.34 & 0.61 & 0.72 &
13.06/30 & 0.63 \\ DDO 185 & 0.674 & -2.37 & 0.90 & 0.21 & 6.04/5 & 0.87 \\
DDO 189 & 0.526 & -1.87 & 0.69 & 3.14 & 0.47/8 & 0.21 \\ UGC 10310
& 0.608 & -1.61 & 0.65 & 1.04 & 3.93/13 & 0.50 \\ \hline
\end{tabular}
\end{center}
\label{tab}
\end{table}
In order to observationally check the above result, we have
considered a sample of LSB galaxies with well measured HI +
H$\alpha$ rotation curves extending far beyond the visible edge of
the system. LSB galaxies are known to be ideal candidates to test
DM models since, because of their high gas content, the
rotation curves can be well measured and corrected for possible
systematic errors by comparing 21\,-\,cm HI line emission with
optical H$\alpha$ and ${\rm [NII]}$ data. Moreover, they are
supposed to be DM dominated so that fitting their
rotation curves without this elusive component is a strong
evidence in favour of any successful alternative theory of gravity.
Our sample contains 15 LSB galaxies with data on both the rotation
curve, the surface mass density of the gas component and
$R$\,-\,band disk photometry extracted from a larger sample
selected by  \citep{dbb02}. We assume the stars are
distributed in an infinitely thin and circularly symmetric disk
with surface density $\Sigma(R) = \Upsilon_\star I_0
exp{(-R/R_d)}$ where the central surface luminosity $I_0$ and the
disk scalelength $R_d$ are obtained from fitting to the stellar
photometry. The gas surface density has been obtained by
interpolating the data over the range probed by HI measurements
and extrapolated outside this range.

When fitting to the theoretical rotation curve, there are three
quantities to be determined, namely the stellar
mass\,-\,to\,-\,light (M/L) ratio, $\Upsilon_{\star}$ and the
theory parameters $(\beta, r_c)$. It is worth stressing that,
while fit results for different galaxies should give the same
$\beta$, $r_c$ must be set on a galaxy\,-\,by\,-\,galaxy basis.
However, it is expected that galaxies having similar properties in
terms of mass distribution have similar values of $r_c$ so that
the scatter in $r_c$ must reflect somewhat that on the terminal
circular velocities. In order to match the model with the data, we
perform a likelihood analyzis determining for each galaxy using as
fitting parameters $\beta$, $\log{r_c}$ (with $r_c$ in kpc) and
the gas mass fraction\footnote{This is related to the $M/L$ ratio
as $\Upsilon_{\star} = [(1 - f_g) M_{g}]/(f_g L_d)$ with $M_g =
1.4 M_{HI}$ the gas (HI + He) mass, $M_d = \Upsilon_{\star} L_d$
and $L_d = 2 \pi I_0 R_d^2$ the disk total mass and luminosity.}
$f_g$. Considering the results summarized in Table \ref{tab},
the experimental data are successfully fitted by the model. In
particular, for the best fit range of $\beta$ $(\beta=0.58\pm
0.15)$, one can conclude that $R^n$ gravity with $1.34 < n <2.41$
(which well overlaps the above mentioned range of $n$ interesting
in cosmology) can be a good candidate to solve the missing matter
problem in LSB galaxies without any DM \citep{cardone}.
\begin{figure}
\centering\resizebox{6cm}{!}{\includegraphics{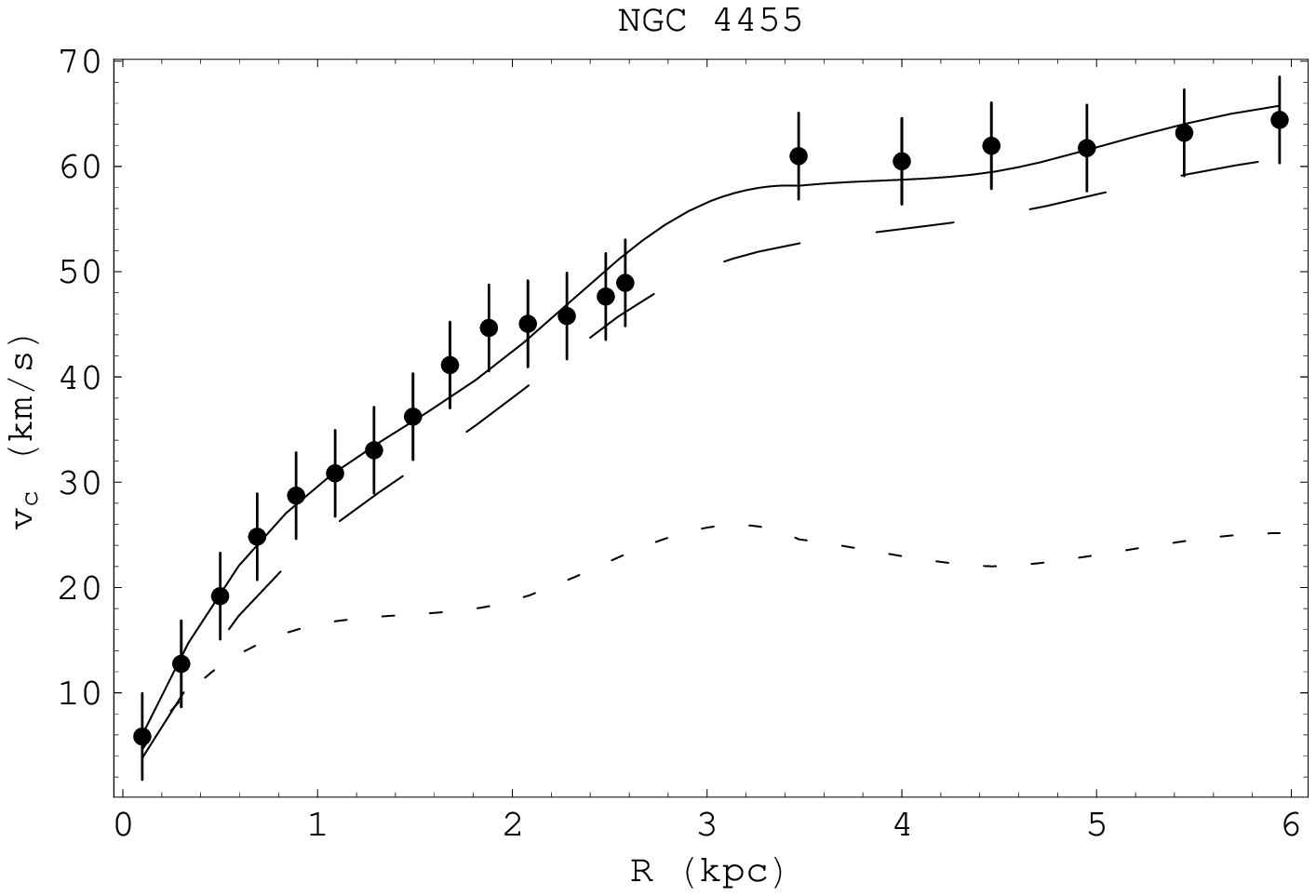}}
\centering\resizebox{6cm}{!}{\includegraphics{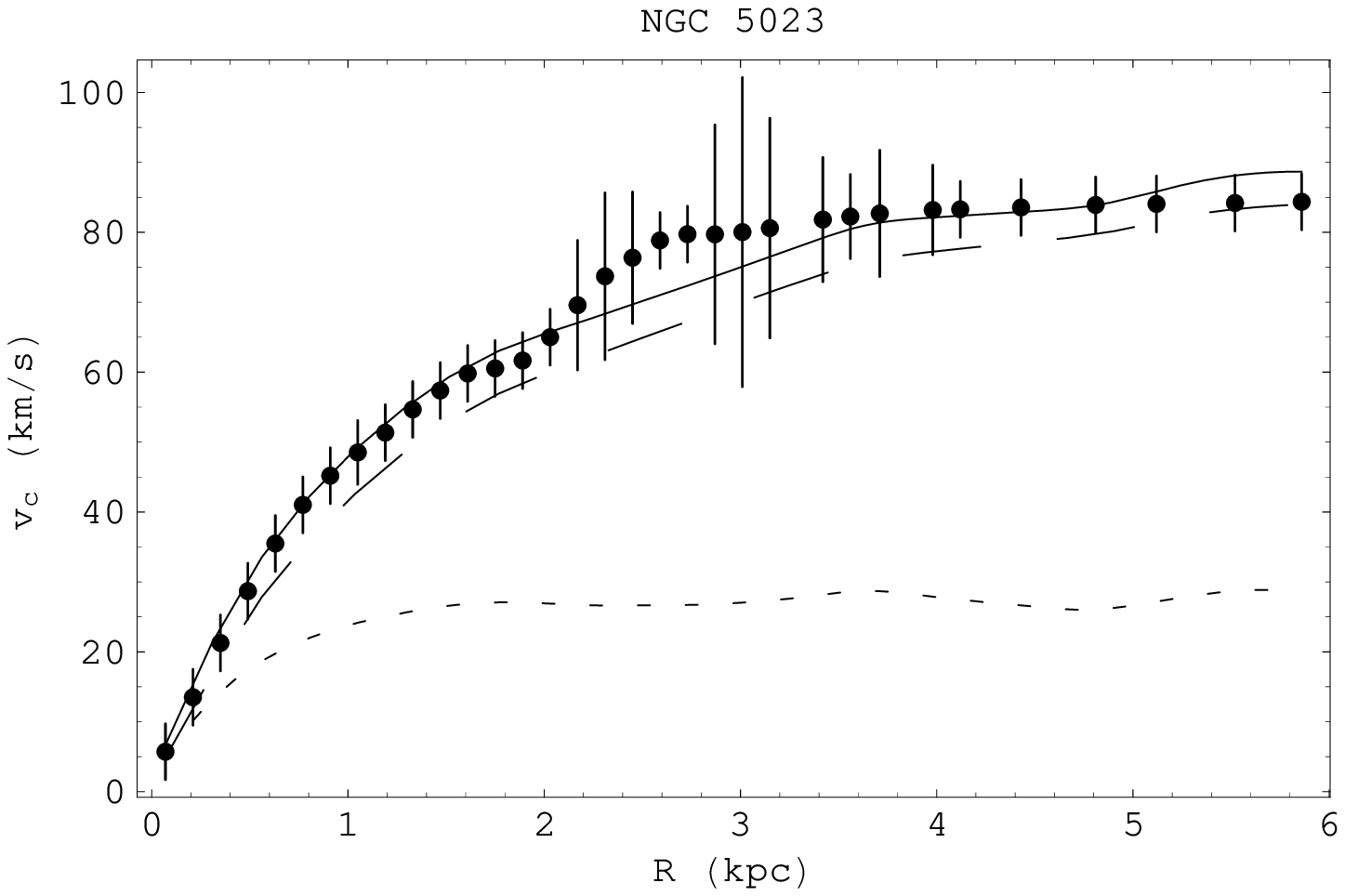}}
\caption{Best fit theoretical rotation curve superimposed to the
data for the LSB galaxy NGC 4455 (left) and NGC 5023 (right). To
better show the effect of the correction to the Newtonian
gravitational potential, we report the total rotation curve
$v_c(R)$ (solid line), the Newtonian one (short dashed) and the
corrected term (long dashed).\label{fig: lsb1}}
\end{figure}
At this point, it is worth wondering whether a link may be found
between $R^n$ gravity and the standard approach based on dark
matter haloes since both theories fit equally well the same data.
 As a matter of fact, it is possible to
define an {\it effective DM halo} by imposing that its
rotation curve equals the correction term to the Newtonian curve
induced by $R^n$ gravity. Mathematically, one can split the total
rotation curve derived from $R^n$ gravity as $v_c^2(r) = v_{c,
N}^2(r) + v_{c, corr}^2(r)$ where the second term is the
correction one. Considering, for simplicity a spherical halo
embedding an infinitely thin exponential disk, we may also write
the total rotation curve as $v_c^2(r) = v_{c, disk}^2(r) + v_{c,
DM}^2(r)$ with $v_{c, disk}^2(r)$ the Newtonian disk rotation
curve and $v_{c, DM}^2(r) = G M_{DM}(r)/r$ the DM one,
$M_{DM}(r)$ being its mass distribution. Equating the two
expressions, we get\,:

\begin{equation}
M_{DM}(\eta) = 2^{\beta - 5} \eta_c^{-\beta} \pi (1 - \beta)
\Sigma_0 R_d^2 \eta^{\frac{\beta + 1}{2}} {\cal{I}}_0(\eta, \beta)
\ . \label{eq: mdm}
\end{equation}
with $\eta = r/R_d$, $\Sigma_0 = \Upsilon_{\star} I_0$ and\,:
\begin{equation} {\cal{I}}_0(\eta, \beta) =
\int_{0}^{\infty}{{\cal{F}}_0(\eta, \eta', \beta) k^{3 - \beta}
\eta'^{\frac{\beta - 1}{2}} {\rm e}^{- \eta'} d\eta'} \label{eq:
deficorr}
\end{equation}
with ${\cal{F}}_0$ only depending on the geometry of the system.
Eq.(\ref{eq: mdm}) defines the mass profile of an effective
spherically symmetric DM halo whose ordinary rotation
curve provides the part of the corrected disk rotation curve due
to the addition of the curvature corrective term to the
gravitational potential. It is evident that, from an observational
viewpoint, there is no way to discriminate between this dark halo
model and $R^n$ gravity. Having assumed spherical symmetry for the
mass distribution, it is immediate to compute the mass density for
the effective dark halo as $\rho_{DM}(r) = (1/4 \pi r^2)
dM_{DM}/dr$. The most interesting features of the density profile
are its asymptotic behaviours that may be quantified by the
logarithmic slope $\alpha_{DM} = d\ln{\rho_{DM}}/d\ln{r}$ which
can be computed only numerically as function of $\eta$ for fixed
values of $\beta$ (or $n$). The asymptotic values at the center
and at infinity denoted as $\alpha_0$ and $\alpha_{\infty}$ result
particularly interesting. It turns out that $\alpha_0$ almost
vanishes so that in the innermost regions the density is
approximately constant. Indeed, $\alpha_0 = 0$ is the value
corresponding to models having an inner core such as the cored
isothermal sphere and the Burkert model \citep{burk}. Moreover, it
is well known that galactic rotation curves are typically best
fitted by cored dark halo models \citep{GS04}. On the
other hand, the outer asymptotic slope is between $-3$ and $-2$,
that are values typical of most dark halo models in literature. In
particular, for $\beta = 0.58$ one finds $(\alpha_0,
\alpha_{\infty}) = (-0.002, -2.41)$, which are quite similar to
the value for the Burkert model $(0, -3)$, that has been
empirically proposed to provide a good fit to the LSB and dwarf
galaxies rotation curves. The values of $(\alpha_0,
\alpha_{\infty})$ we find for our best fit effective dark halo
therefore suggest a possible theoretical motivation for the
Burkert\,-\,like models. Now, due to the construction, the
properties of the effective DM halo are closely related
to the disk one. As such, we do expect some correlation between
the dark halo and the disk parameters. To this aim, exploiting the
relation between the virial mass and the disk parameters , one can
obtain a relation for the Newtonian virial velocity $V_{vir} = G
M_{vir}/R_{vir}$\,:

\begin{equation}
M_d = \frac{(3/4 \pi \delta_{th} \Omega_m \rho_{crit})^{\frac{1 -
\beta}{4}} R_d^{\frac{1 + \beta}{2}} \eta_c^{\beta}}{2^{\beta - 6}
 (1 - \beta) G^{\frac{5 - \beta}{4}}} \frac{V_{vir}^{\frac{5 -
\beta}{2}}}{{\cal{I}}_0(V_{vir}, \beta)} \label{eq: btfvir} \ .
\end{equation}
We have numerically checked that Eq.(\ref{eq: btfvir}) may be well
approximated as $M_d \propto V_{vir}^{a}$ which has the same
formal structure as the baryonic Tully\,-\,Fisher (BTF) relation
$M_b \propto V_{flat}^a$ with $M_b$ the total (gas + stars)
baryonic mass and $V_{flat}$ the circular velocity on the flat
part of the observed rotation curve. In order to test whether the
BTF can be explained thanks to the effective DM halo we
are proposing, we should look for a relation between $V_{vir}$ and
$V_{flat}$. This is not analytically possible since the estimate
of $V_{flat}$ depends on the peculiarities of the observed
rotation curve such as how far it extends and the uncertainties
on the outermost points. For given values of the disk parameters,
we therefore simulate theoretical rotation curves for some values
of $r_c$ and measure $V_{flat}$ finally choosing the fiducial
value for $r_c$ that gives a value of $V_{flat}$ as similar as
possible to the measured one. Inserting the relation thus found
between $V_{flat}$ and $V_{vir}$ into Eq.(\ref{eq: btfvir}) and
averaging over different simulations, we finally get\,:
%\begin{equation}
$\log{M_b} = (2.88 \pm 0.04) \log{V_{flat}} + (4.14 \pm 0.09)$
%\label{eq: btfour}
%\end{equation} while a direct fit to the
while observational data give\,:
%\begin{equation}
$\log{M_b} = (2.98 \pm 0.29) \log{V_{flat}} + (3.37 \pm 0.13) \ .$
%\label{eq: btfssm}
%\end{equation}
The slope of the predicted and observed BTF are in good agreement
thus leading further support to our approach. The zeropoint is
markedly different with the predicted one being significantly
larger than the observed one, but it is worth stressing, however,
that both relations fit the data with similar scatter. A
discrepancy in the zeropoint may be due to our approximate
treatment of the effective halo which does not take into account
the gas component. Neglecting this term, we should increase the
effective halo mass and hence $V_{vir}$ which affects the relation
with $V_{flat}$ leading to a higher than observed zeropoint.
Indeed, the larger is $M_g/M_d$, the more the point deviate from
our predicted BTF thus confirming our hypothesis. Given this
caveat, we may therefore conclude with confidence that $R^n$
gravity offers a theoretical foundation even
for the empirically found BTF relation. \\
These results are referred to a simple
choice of $f(R)$, while it is likely that a more complicated
Lagrangian is needed to reproduce the whole dark sector
phenomenology at all scales. Nevertheless, although not
definitive, these achievements represent an intriguing matter for
future more exhaustive investigations. In particular, exploiting
such models can reveal a useful approach to motivate a more
careful search for a single fundamental theory of gravity able to
explain the full cosmic dynamics with the only two ingredients we
can directly experience, namely the background gravity and the
baryonic matter \citep{jcap}.

\section{Conclusions}

DM and DE can be considered among the biggest puzzles of modern physics. While their  effects are evident from galactic to cosmological scales, their detection, at fundamental level, is a cumbersome task that, up to now, has no  definitive answer. From a macroscopic viewpoint, DM is related to the clustering of astrophysical  structures and DE is an unclustered form of energy (e.g. the cosmological constant $\Lambda$) which should be the source of cosmic speed up, detected by SNeIa and other cosmological indicators in the Hubble flow. 
Following the standard approach in physics, such {\it missing matter problem} and {\it cosmic acceleration} should be related to some fundamental ingredient (e.g. particles). Specifically, such particles should be not electromagnetically  interacting (and then they are "dark") and,  a part a few percent of  further baryons, should be out of SM. In the case of DE, instead, they should be scalar fields that do not  give rise to clustered structures. 

Essentially, the DM (and DE) problem consists of three issues:
$i)$ the existence, $ii)$ the detection, $iii)$  the possible alternatives. In this review paper, we have tried to  summarize the status of art with no claim to  completeness.

After a discussion of the need of DM to address astrophysical and cosmological dynamics, we have reviewed the  possible candidates. as SUSY, WIMP and extra dimension particles. 
The big issue is the detection of such particles that can be faced in three ways. 

The direct detection (e.g. of WIMPs) consists, essentially, in
identifying a certain mass $m_X$ from scattering processes.  Several
underground experiments (e.g. DAMA) are running or have to start with
this purpose. The approach is that standard interacting particles have
to be shielded by some layer of rocks so then the weak interacting
particles could be identified.

Indirect searches look for the excesses of annihilation products that
should be DM remnants. Such experiments, like PAMELA and FERMI, are
based on spacecrafts and are devoted to reconstruct the signature of
DM.

Finally, DM evidences could come from particle colliders like LHC or
Tevatron. In this case, DM should result from high-energy colliding
processes where particles out of SM should be produced.

A radically alternative view is that DM (and DE) do not exist at all.
Missing matter problem and cosmic speed up could be addressed by
revising and extending the gravitational sector. Specifically,
Einstein's GR should be a theory working only at local scale that
should be revised at infrared (large scale structure and cosmology)
and ultraviolet regimes (quantum gravity).

In conclusion, it seems that DM problem will be open until incontrovertible evidences (signatures) will be available at fundamental level. Very likely, the today working space- and ground-based experiments  are going to reach the requested energy and precision levels in order to confirm or rule out the DM issue.

\section{Acknowledgement}
The authors thank G Barbarino for discussions and comments on
the topics of this review paper.

\end{document}